\documentclass[sigconf,authorversion]{acmart}

\usepackage{amsmath}
\usepackage{graphicx}
\usepackage{xcolor}

\copyrightyear{2022} 
\acmYear{2022} 
\setcopyright{acmlicensed}\acmConference[SA '22 Technical Communications]{SIGGRAPH Asia 2022 Technical Communications}{December 6--9, 2022}{Daegu, Republic of Korea}
\acmBooktitle{SIGGRAPH Asia 2022 Technical Communications (SA '22 Technical Communications), December 6--9, 2022, Daegu, Republic of Korea}
\acmPrice{15.00}
\acmDOI{10.1145/3550340.3564231}
\acmISBN{978-1-4503-9465-9/22/12}

\copyrightyear{2022} 
\acmYear{2022} 
\setcopyright{acmlicensed}\acmConference[SA '22 Technical Communications]{SIGGRAPH Asia 2022 Technical Communications}{December 6--9, 2022}{Daegu, Republic of Korea}
\acmBooktitle{SIGGRAPH Asia 2022 Technical Communications (SA '22 Technical Communications), December 6--9, 2022, Daegu, Republic of Korea}
\acmPrice{15.00}
\acmDOI{10.1145/3550340.3564231}
\acmISBN{978-1-4503-9465-9/22/12}

\setcitestyle{square}

\begin{document}

\title{Combining GPU Tracing Methods within a Single Ray Query}

\author{Pieterjan Bartels}
\affiliation{%
  \institution{Advanced Micro Devices, Inc.}
  \city{Brussels}
  \country{Belgium}}
\email{Pieterjan.Bartels@amd.com}
\orcid{0000-0001-5342-4841}

\author{Takahiro Harada}
\affiliation{%
  \institution{Advanced Micro Devices, Inc.}
  \city{Santa Clara}
  \country{USA}}
\email{Takahiro.Harada@amd.com}
\orcid{0000-0001-5158-8455}
  
\renewcommand{\shortauthors}{Bartels and Harada}

\begin{abstract}
A recent trend in real-time rendering is the utilization of the new hardware ray tracing capabilities. Often, usage of a distance field representation is proposed as an alternative when hardware ray tracing is deemed too costly, and the two are seen as competing approaches. In this work, we show that both approaches can work together effectively for a single ray query on modern hardware. We choose to use hardware ray tracing where precision is most important, while avoiding its heavy cost by using a distance field when possible. While a simple approach, in our experiments the resulting tracing algorithm overcomes the associated overhead and allows a user-defined middle ground between the performance of distance field traversal and the improved visual quality of hardware ray tracing. 
\end{abstract}




\begin{teaserfigure}
  \begin{tabular}{c c c c c c}
  \multicolumn{2}{c}{\textbf{HWRT only}} &
  \multicolumn{2}{c}{\textbf{Combined (Ours)}} &
  \multicolumn{2}{c}{\textbf{DF Only}} \\
  \multicolumn{2}{c}{\includegraphics[trim={20cm 0 5cm 0},clip,width=0.32\textwidth]{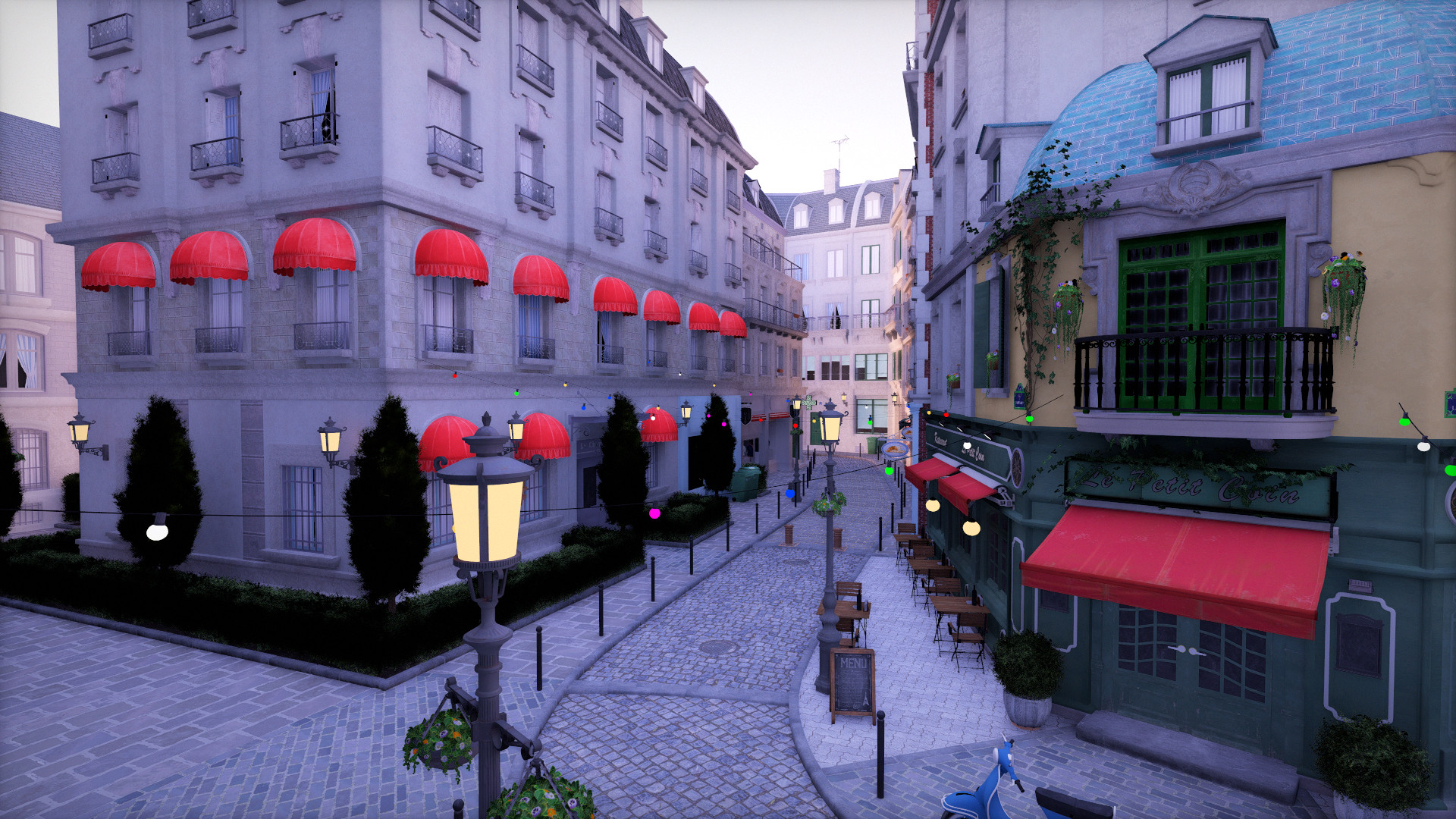}} &
  \multicolumn{2}{c}{\includegraphics[trim={20cm 0 5cm 0},clip,width=0.32\textwidth]{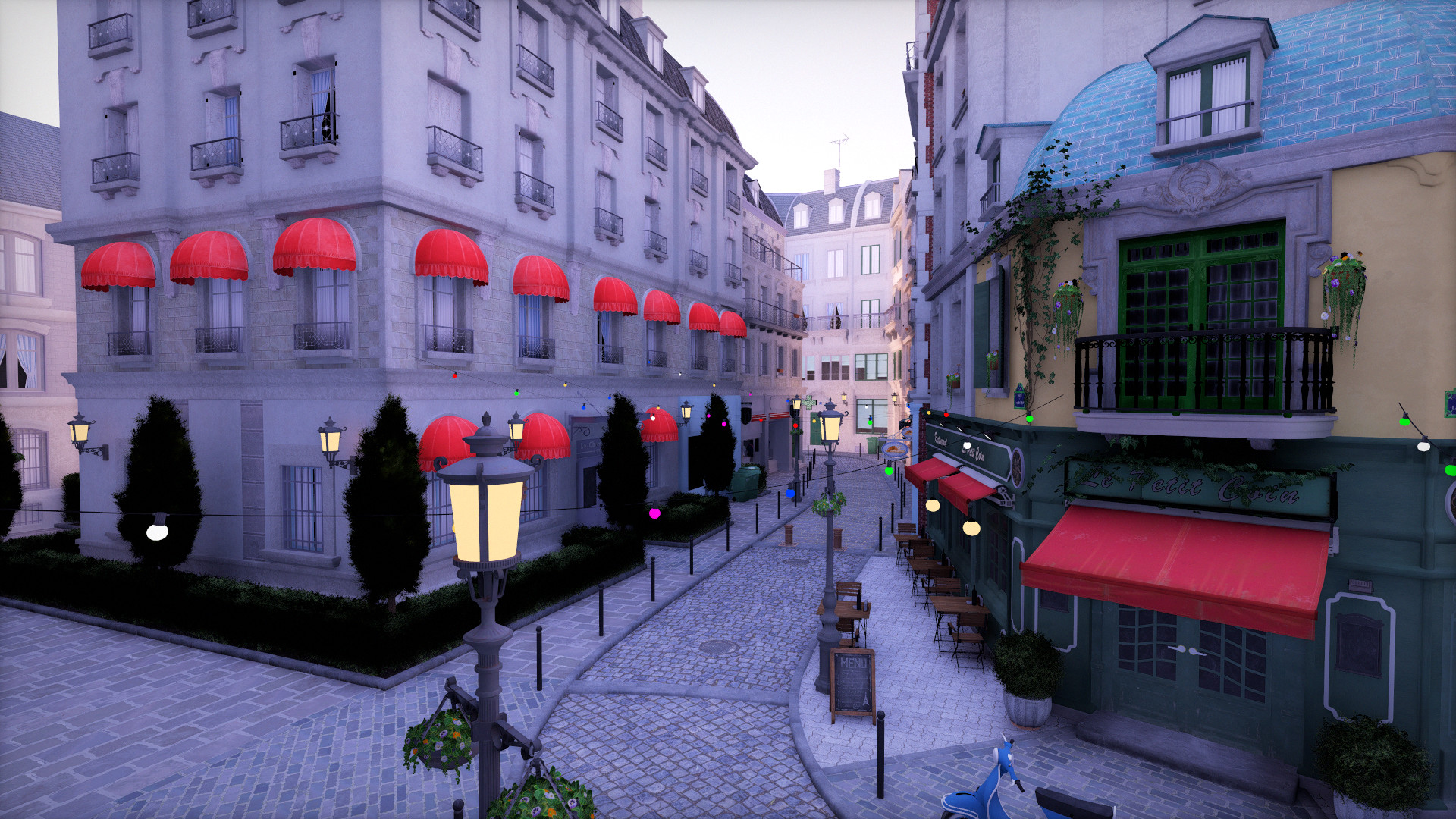}} &
  \multicolumn{2}{c}{\includegraphics[trim={20cm 0 5cm 0},clip,width=0.32\textwidth]{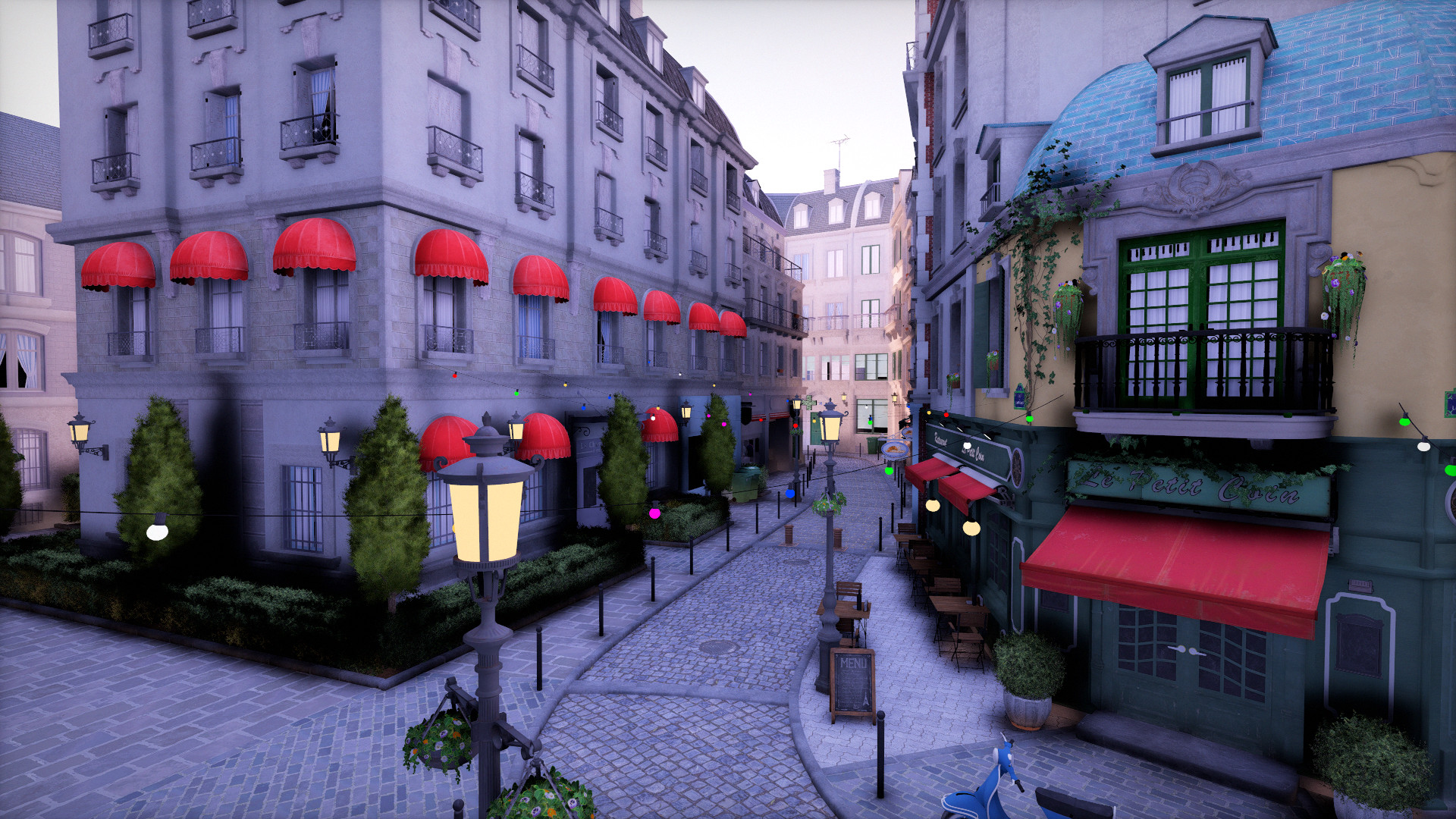}} \\
   \multicolumn{1}{l}{GI-1.0 Time:} & \multicolumn{1}{r}{5.66 ms} & & \multicolumn{1}{r}{4.55 ms} & &   \multicolumn{1}{r}{3.58 ms}\\
  \end{tabular}
  \caption{Our combined tracing method applied within GI-1.0 \cite{gi10}, shown with the per-frame time of the GI algorithm. Notice that with our method, we do not have the obvious artefacts in corners and in shadows. These artefacts are due to inaccuracies in the distance field, which we avoid, while saving 1.2 ms compared to the hardware ray tracing only solution. While not visually identical, our result is much closer and more perceptually pleasing without paying the full cost of hardware ray tracing. Full resolution images can be found in our supplemental material.}
   \label{gi10_results}
\end{teaserfigure}
\maketitle
\section{Introduction \& Related Work}

Ray tracing is a commonly used technique in both offline and real-time rendering algorithms \cite{rtr}. While offline rendering has seen adoption of ray tracing-based techniques for many years, real-time rendering has not been able to resolve many ray queries per frame until the relatively recent introduction of hardware ray tracing (HWRT), allowing the traversal of a ray tracing acceleration structure on GPU \cite{hwrt}.
Unfortunately, within the context of a commercial game engine, HWRT is still too costly to be adopted without constraints, and these engines often have a maximum number of ray queries they can send out per frame, e.g. one query per pixel \cite{hybridRendering}. Introduction of HWRT thus becomes a complex balancing exercise where engineers need to maintain the required frame-rates by designing algorithms that use small numbers of HWRT queries.
When a rendering technique requires more ray queries, or the target platform does not support HWRT, many authors propose tracing against a distance field representation of the scene instead \cite{dfExample}. While this is much faster, artefacts and inaccuracies are introduced as we now no longer trace against the true scene surface.


In this work, we propose the use of both techniques together to handle a single ray query, allowing more efficient resolution of queries within a given budget. To the best of our knowledge we are the first to combine HWRT and distance field tracing in this way. In short, our contributions are as follows:
\begin{itemize}
\item We show that maintaining two data structures and breaking one query into several queries can be worthwhile. 
\item We show that the obvious artefacts introduced by using a lightweight distance field representation can be alleviated by careful selection of the parts of the ray that can instead be hardware ray traced.
\end{itemize}

It is common to combine HWRT with screen-space tracing to improve performance, which is similar to our technique \cite{ss}. However, screen-space tracing is dependent on what is visible in the frame and is usually used to trace the beginning of rays. Our technique has no dependence on the frame whatsoever and allows HWRT for any part of the ray query. Finally, some authors try to improve the performance of ray tracing by avoiding ray divergence through ray sorting \cite{meister2020ray}. This research direction is orthogonal to our work.

\section{Concept}

Our basic principle is very straightforward. We are interested in processing ray queries of the following form:
\begin{equation}
    (o, d, t_{min}, t_{max}, f).
\end{equation}

That is, we are interested in finding out whether there are intersections along a ray $r$ with origin $o$ and normalized direction $d$, that is $r = o + t * d$, where $t$ is used to indicate distance along the ray relative to the origin. The parameters $t_{min}$ and $t_{max}$ place a constraint on the query to only take into account intersections that have a distance $t$ from the origin within the interval $[t_{min}, t_{max}]$. In principle, both $t_{min}$ and $t_{max}$ can be within the interval (-inf, +inf), although in practice they are usually positive. 
In what follows we assume, without loss of generality, that $t_{min} \leq t_{max}$. Finally, the $f$ parameter is a flag indicating whether the query needs to find the intersection within the interval that is closest to the origin, or if it suffices to find any intersection within the interval.

We propose to break up the query by dividing the interval $[t_{min}$, $t_{max}]$ into several, mutually exclusive, intervals. For example, for three sub-queries, we would process the following sub-queries:
\begin{equation}
    (o, d, t_{min}, t_{1}, f),
    (o, d, t_{1}, t_{2}, f),
    (o, d, t_{2}, t_{max}, f).
\end{equation}
We can then process each sub-query separately with a different tracing technique. While the idea itself is simple, this would normally not be implemented as the overhead of using several tracing calls would be deemed too costly. We further optimize our technique based on the $f$ flag: if any hit suffices, sub-queries can be processed out of order, starting with the parts traced against the distance field. If the first hit is needed, we process sub-queries in order and avoid remaining sub-queries once an intersection is found.

\paragraph{Determination of $t_i$ \& tracing technique.} 
The key issue is how to determine where to use each technique, i.e. how to determine the intervals $[t_i, t_{i+1}]$. The objective is to use each tracing technique to its strengths: we want to use HWRT where we require high precision (or alpha testing, for example), and  distance field tracing where we want fast traversal but precision is less important. High performance is achieved by limiting HWRT to short and low-divergent ray queries, which depends on the context in which ray tracing is used. In the next section we provide two examples of determining such intervals for specific use cases.

\paragraph{Surface information.} 
The information returned from a ray tracing query depends on the specific needs of the algorithm. An important consideration for our technique is whether the necessary surface information is available in both tracing techniques. Notably, retrieving material information from a distance field is non-trivial.
\begin{figure*} 
  \begin{tabular}{c  c  c  c  c  c}
  \multicolumn{2}{c}{\textbf{HWRT only}} &
  \multicolumn{2}{c}{\textbf{Combined (Ours)}} &
  \multicolumn{2}{c}{\textbf{DF Only}} \\
  \multicolumn{2}{c}{\includegraphics[width=0.31\textwidth]{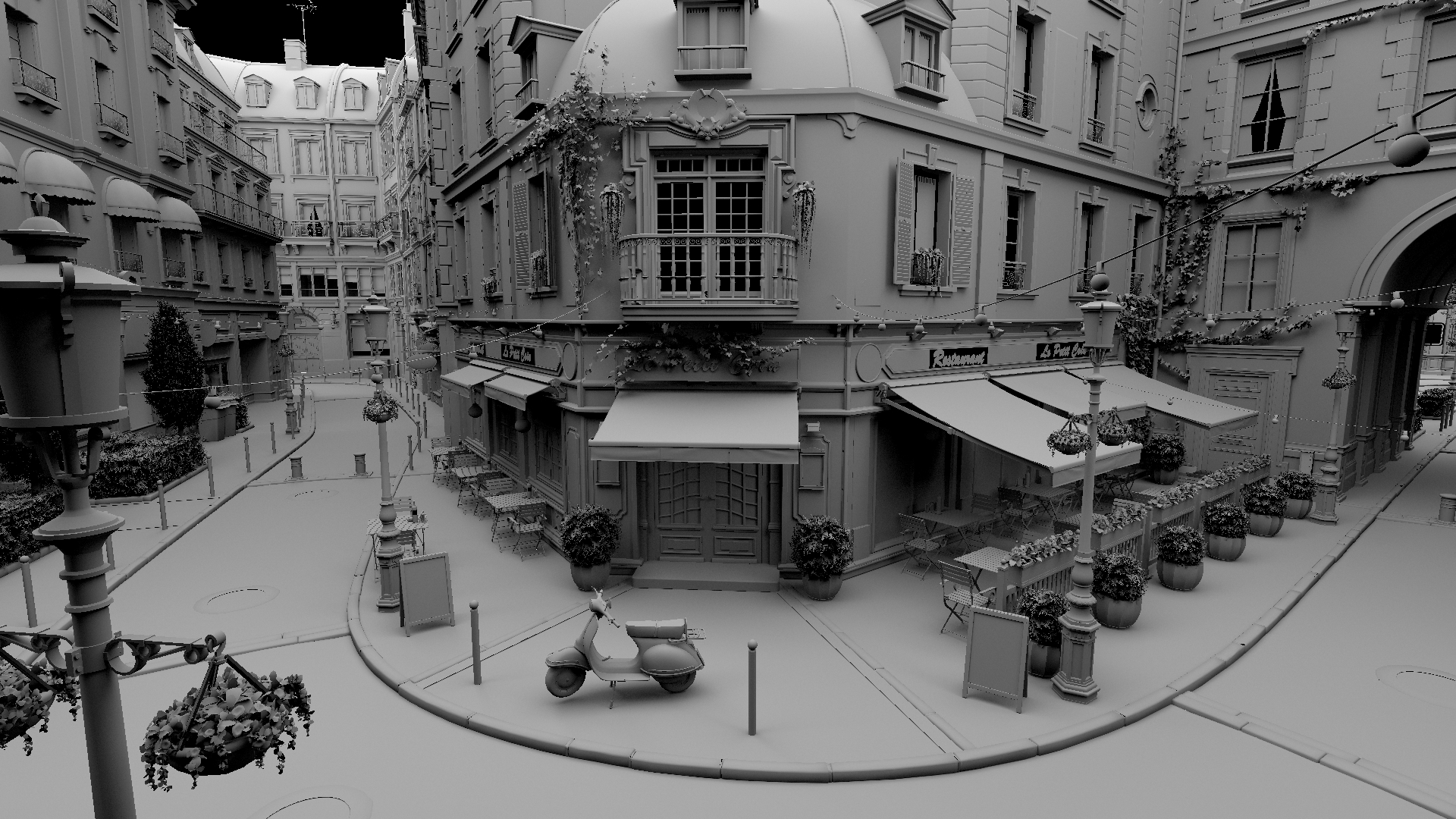}} &
  \multicolumn{2}{c}{\includegraphics[width=0.31\textwidth]{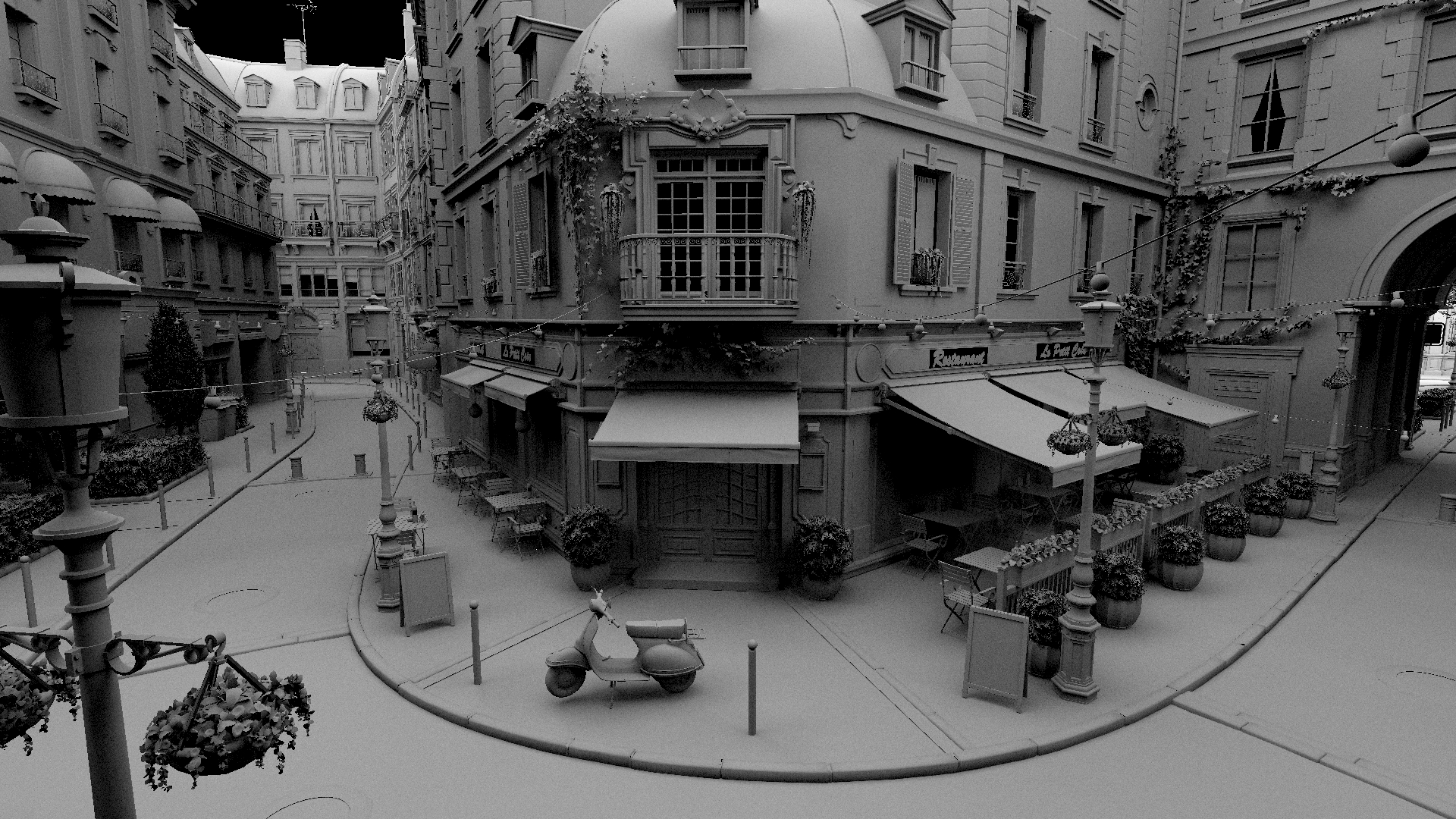}} &
  \multicolumn{2}{c}{\includegraphics[width=0.31\textwidth]{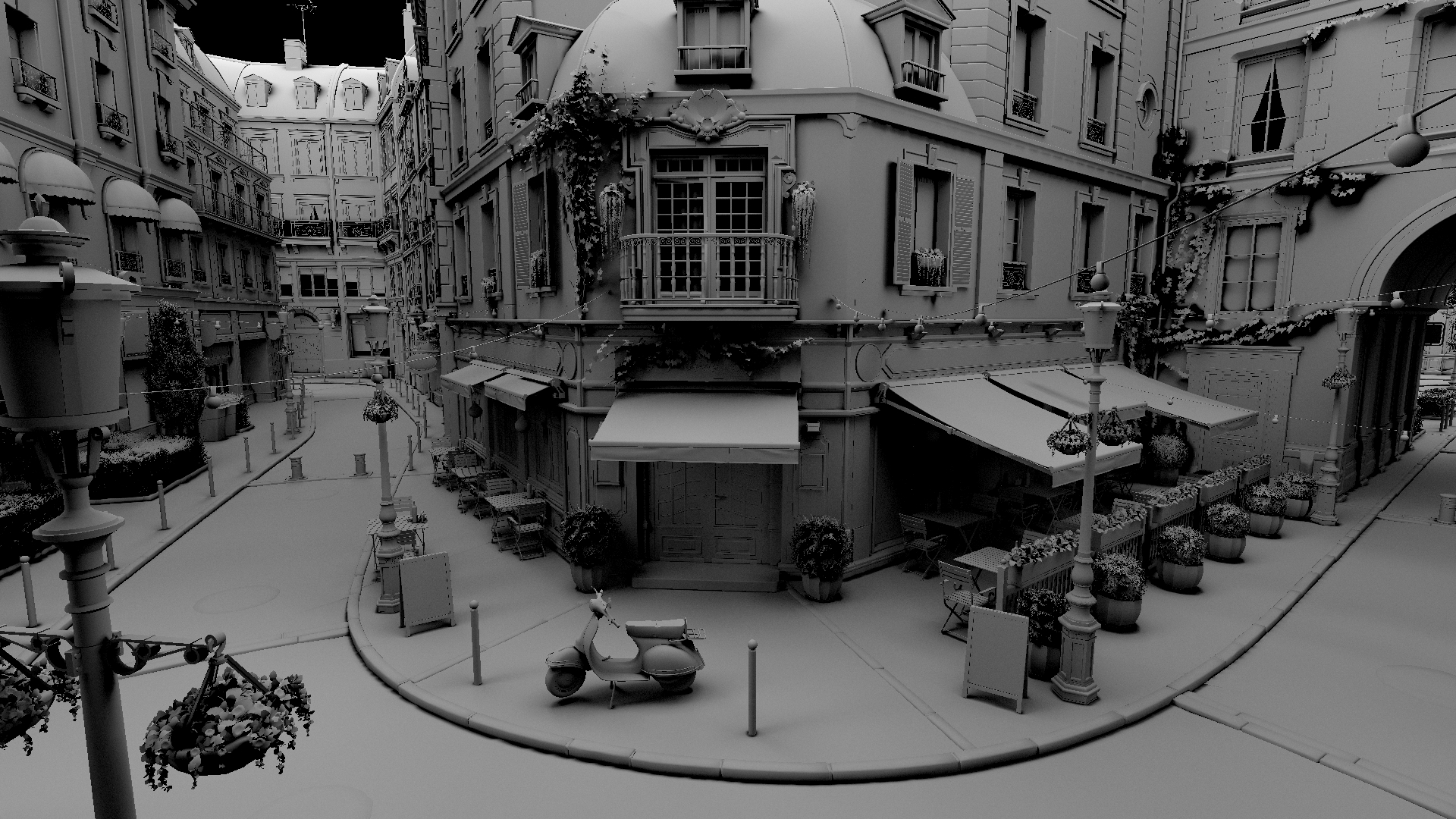}}\\
  Tracing Time: & \multicolumn{1}{r}{8.33 ms} & & \multicolumn{1}{r}{3.16 ms} & & \multicolumn{1}{r}{1.35 ms}\\

  \multicolumn{2}{c}{\includegraphics[width=0.31\textwidth]{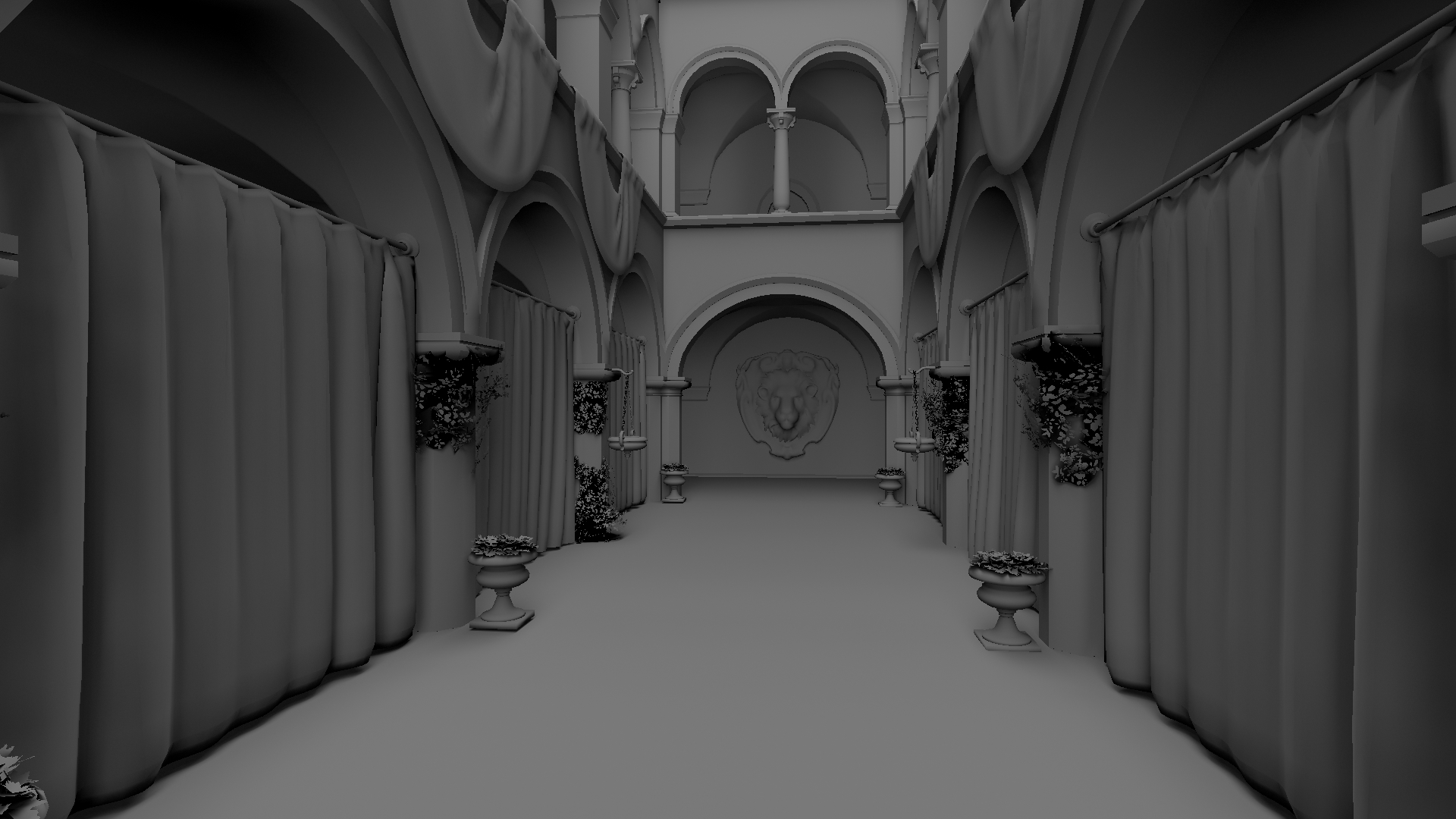}} &
  \multicolumn{2}{c}{\includegraphics[width=0.31\textwidth]{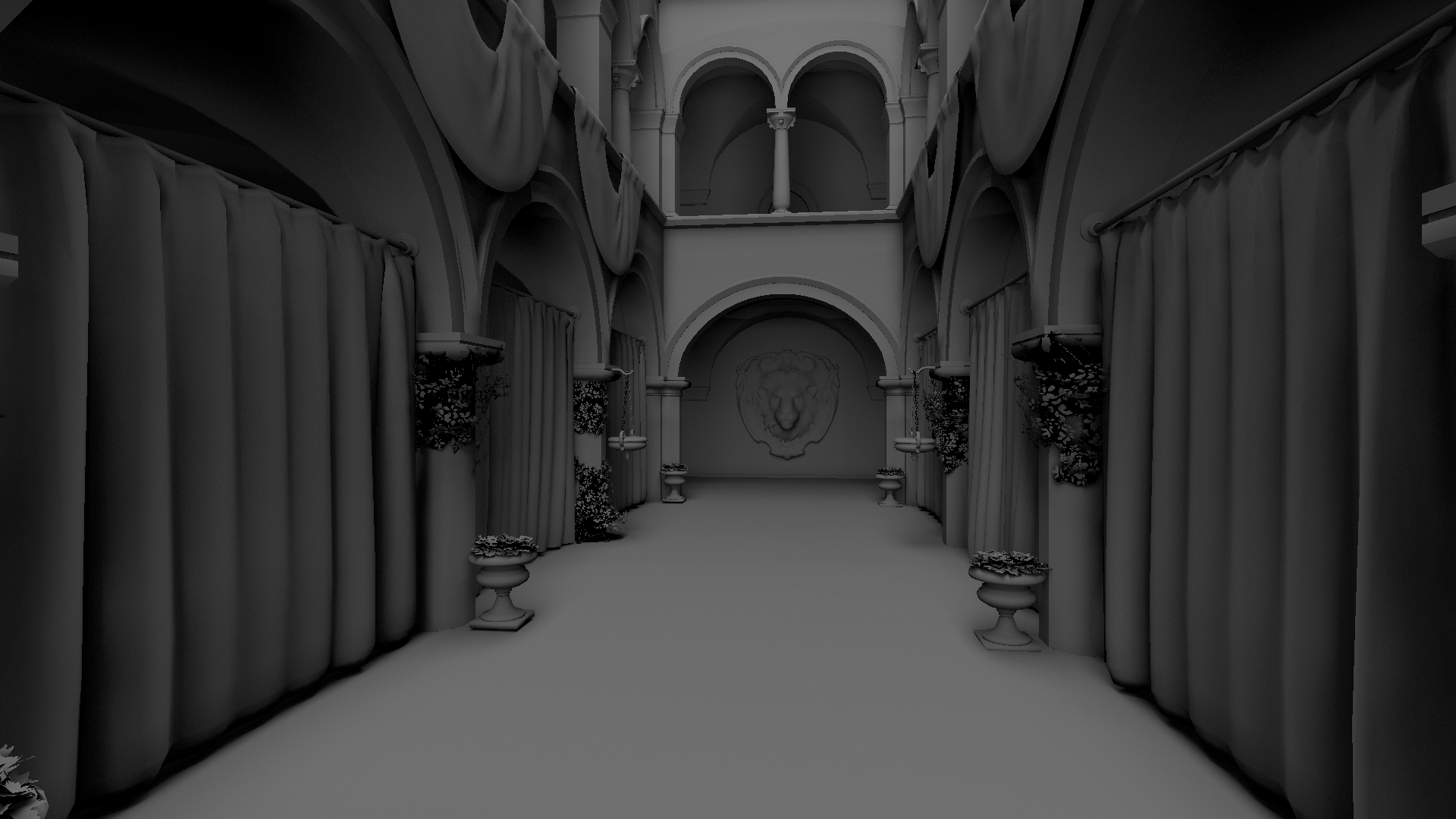}} &
  \multicolumn{2}{c}{\includegraphics[width=0.31\textwidth]{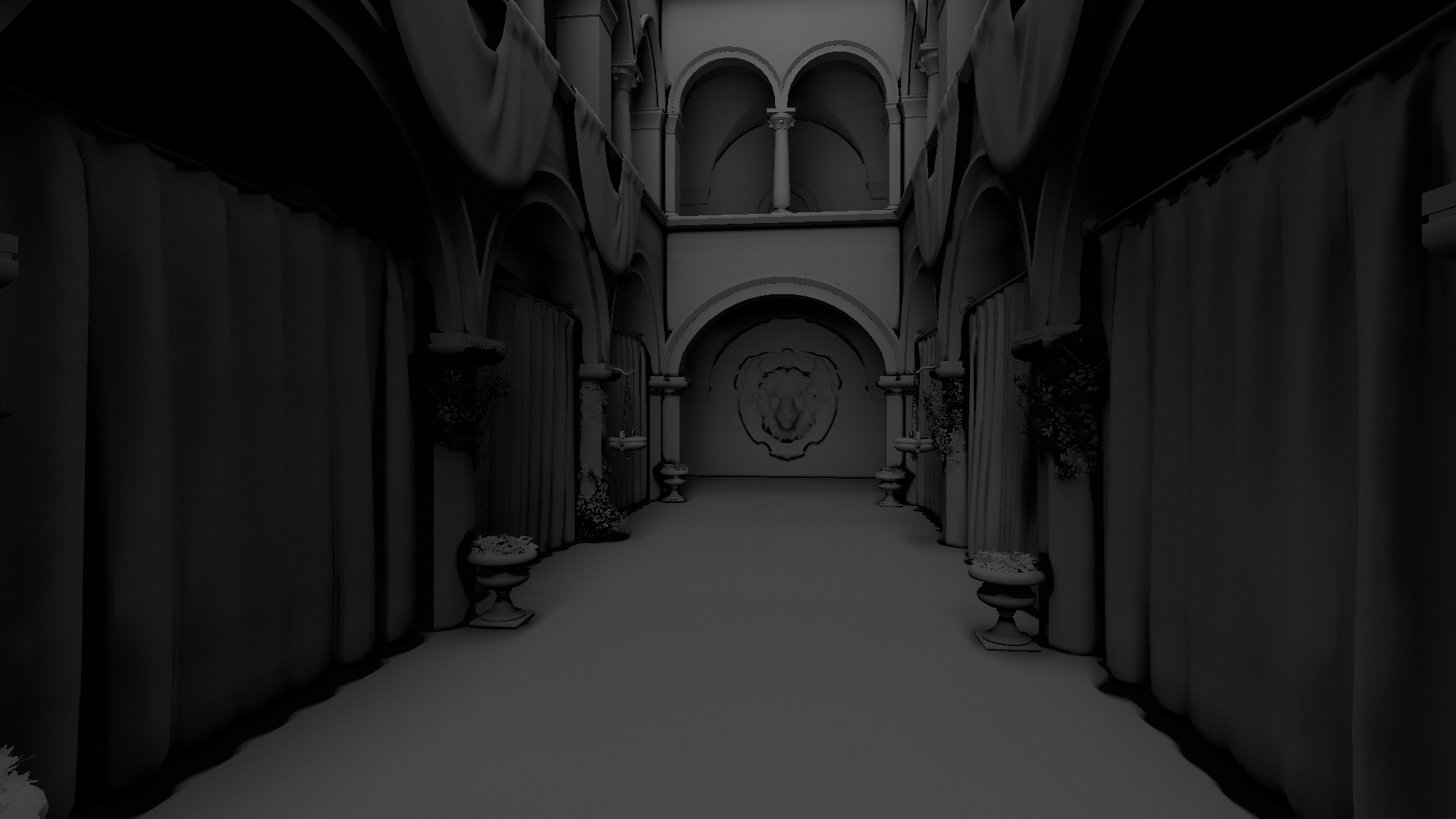}} \\
    Tracing Time: & \multicolumn{1}{r}{2.73 ms} & & \multicolumn{1}{r}{1.86 ms} & & \multicolumn{1}{r}{1.13 ms}\\
  \multicolumn{2}{c}{\includegraphics[width=0.31\textwidth]{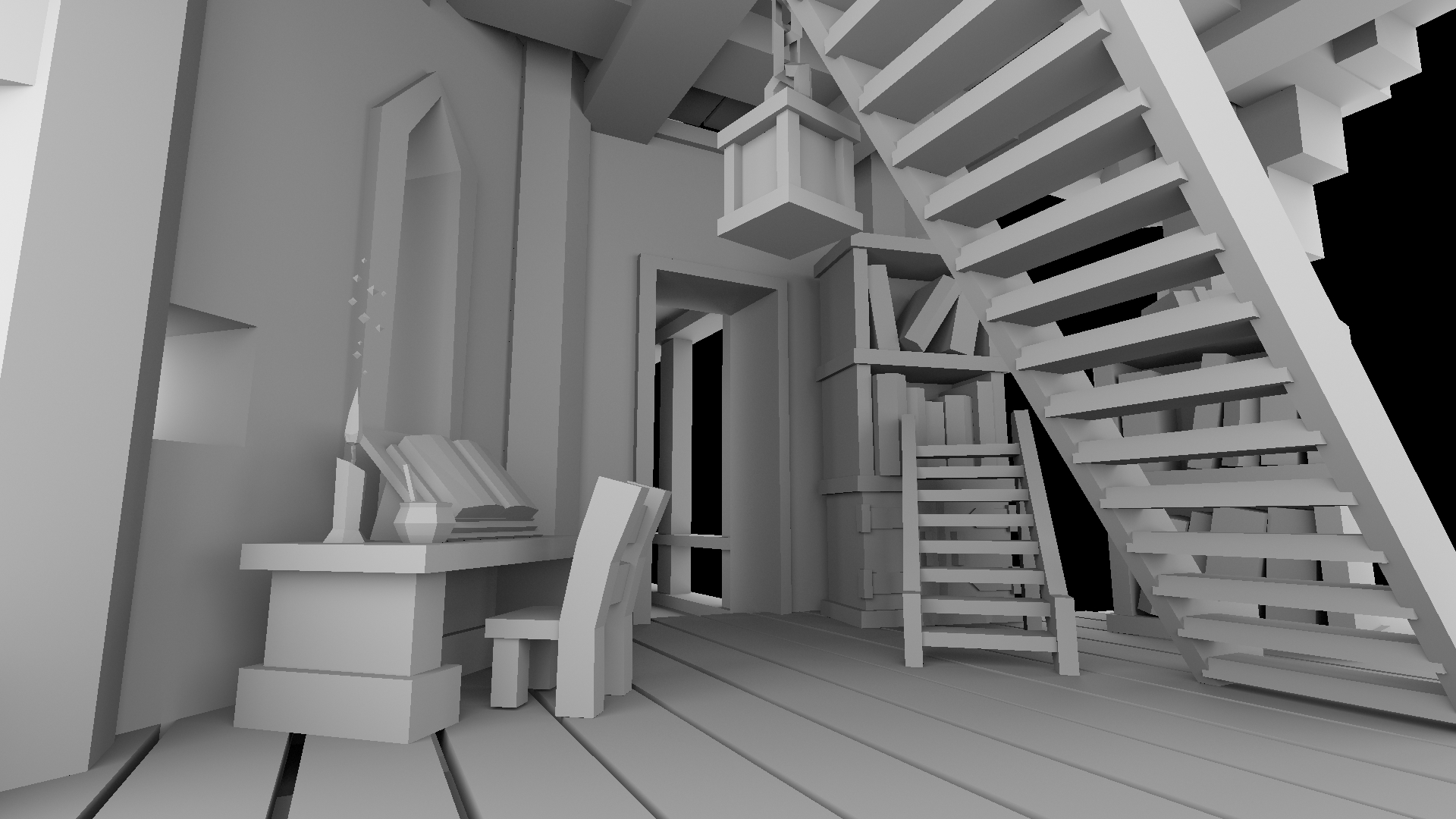}} &
  \multicolumn{2}{c}{\includegraphics[width=0.31\textwidth]{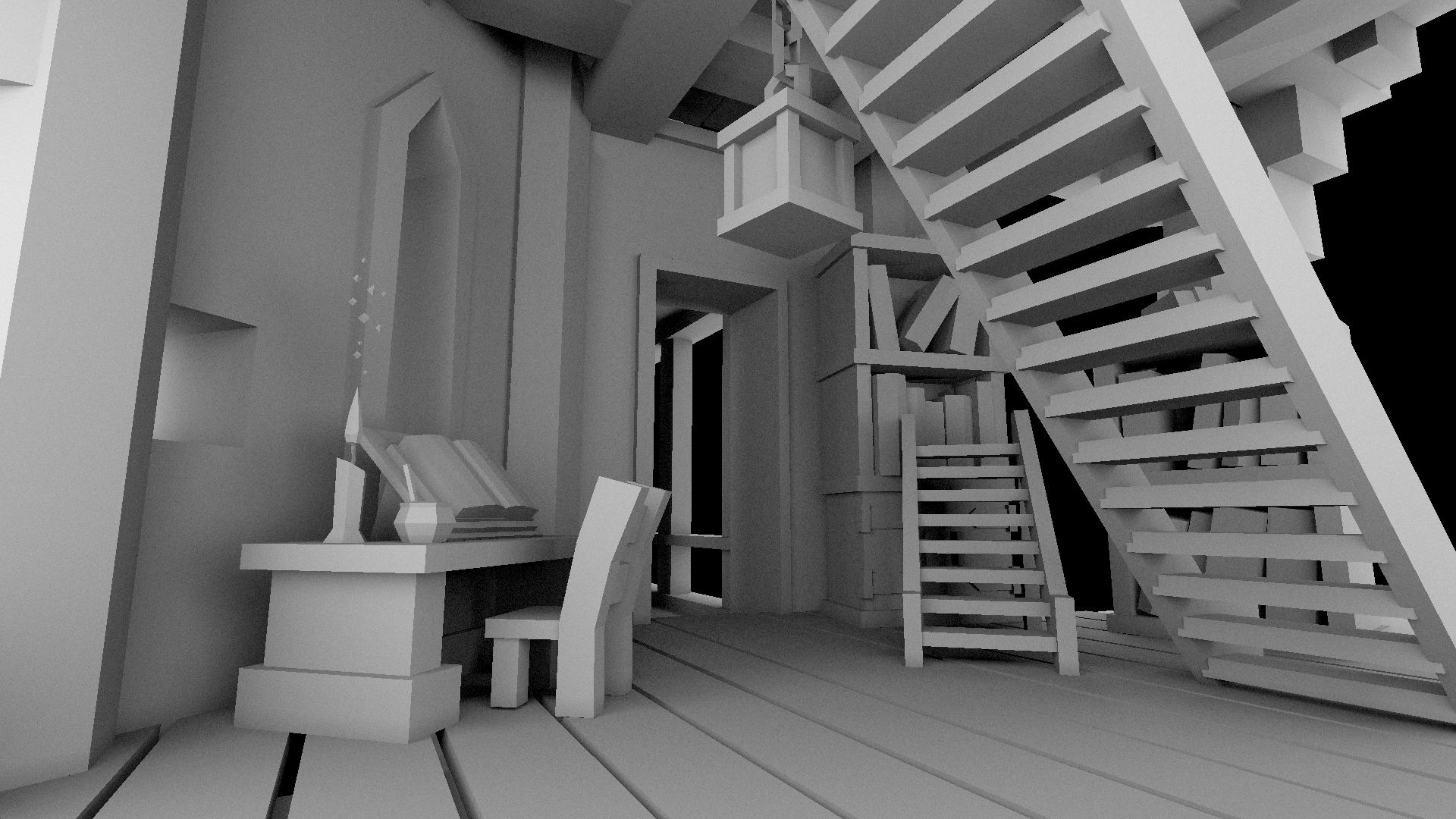}} &
  \multicolumn{2}{c}{\includegraphics[width=0.31\textwidth]{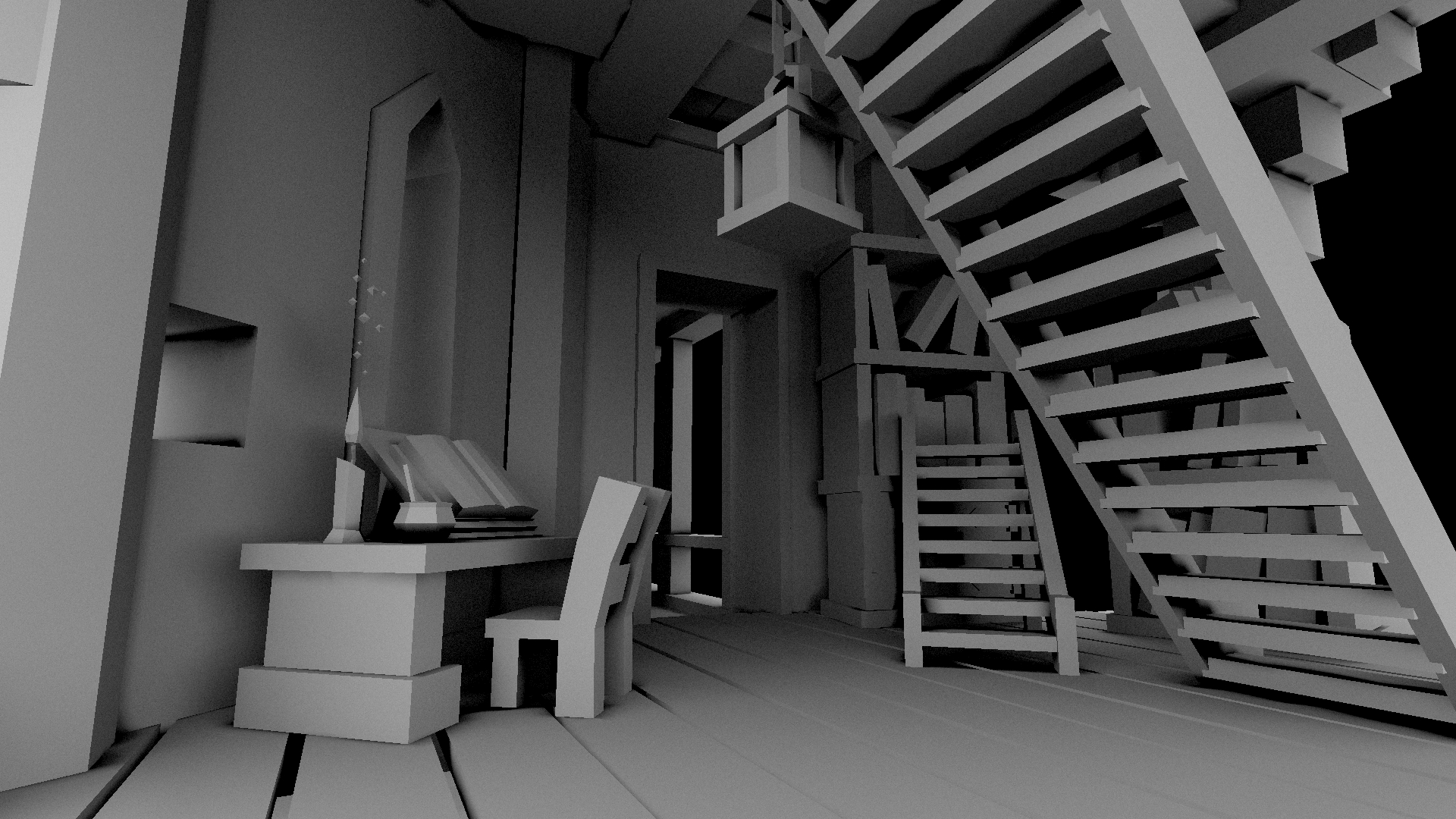}}\\
    Tracing Time: & \multicolumn{1}{r}{1.72 ms} & & \multicolumn{1}{r}{1.64 ms} & & \multicolumn{1}{r}{1.14 ms}\\
  \end{tabular}
  \caption{Our combined tracing method applied to a basic ray traced Ambient Occlusion calculation. In the top row, we see the Bistro scene where the AO max distance is set to 100 for testing purposes. Notice that our result avoids the obvious darkening artefacts from \textit{blobbing} within the distance field, while being traced twice as fast as with hardware ray tracing only. When we render simpler scenes with shorter AO distances such as in the bottom two rows (max AO distance: 10), we maintain consistent results: faster tracing without problematic artefacts. Full resolution and error images can be found in our supplemental material.}
  \label{ao}
\end{figure*}
\section{Experiments \& Results}

In this section, we apply our concept to two types of very typical ray tracing queries found in common rendering algorithms: Occlusion queries and Shadow rays. We run our experiments on an AMD Radeon RX 6800 XT, with the algorithms implemented in a DX12/DXR-based research rendering framework. In both cases, the ray count is proportional to the pixel count, where the ray origin for each pixel is based on an initial rasterization of the scene. Shader invocations are tiled across the viewport image and ray directions are randomized or otherwise divergent within each tile, a very common set-up for stochastic ray tracing algorithms. For both algorithms, we compare our combined tracing method against the \textit{pure} techniques, where all overhead is circumvented.


\paragraph{Distance Field Generation} In both algorithms described below, we prepend a pass to the frame that builds a distance field representation of the scene. The representation is computed through a rasterization-based voxelization algorithm based on the work by Pantaleoni \cite{voxelpipe}, after which the voxelization is converted to a distance field using both the jump flooding algorithm and the fast sweeping method for optimization. The distance field is built in cascades centered around the camera position, to allow for smaller, more accurate voxel-sizes close to the camera. We use this distance field in two different set-ups. When we perform combined tracing, we tune the cascades to a lower resolution: in this case we expect accuracy from HWRT. In practice, our smallest voxel-size is set to 0.1 m in this case, with 5 cascades of 64 voxels in each dimension. This also keeps the memory and performance overhead related to the distance field building and storage smaller in the combined case. When we trace against the distance field only, we double the resolution (i.e. halve the voxel-size) while keeping the same number of cascades, resulting in a larger overhead but more accurate results. Despite this, our results are more visually accurate because of our sparse use of HWRT, as we are going to show below.

Finally, the distance field has support for incremental updates: we do not update the distance field when everything remains static. When the camera moves, we use a rolling scheme to update the changed cascade slices only, which never takes more than 0.1 ms when movement is limited to one slice per frame in each dimension, even on the Bistro scene (Figure \ref{ao}). When the scene is dynamic, we update the whole distance field every frame which takes between 0.5 and 2 ms, depending on the scene and distance field resolution. This scheme amortizes the overhead of building a second data-structure beside the BVH as much as possible.

\paragraph{Occlusion Queries.} We implement a ray tracing based ambient occlusion (AO) algorithm. For each pixel, we send an occlusion ray from the point visible in the frame in a randomized direction to a user-defined maximum AO distance $t_{ao}$ (i.e. $t \in [0, t_{ao}]$). The AO values are accumulated over time through temporal reprojection.

To apply our technique, we need to determine appropriate $t_i$. For ambient occlusion, we want the high precision of HWRT close to the origin, where missing high frequency details can lead to objectionable artefacts. On the other hand, inaccuracies at a greater distance get blurred and can therefore be traced against the distance field. Thus, we define a single value $t_{1}$, resulting in two intervals $[0, t_{1}]$ and $[t_{1}, t_{ao}]$, tracing the first with HWRT. Since intersection distance is typically used in AO calculation, we use first-hit tracing. We compute $t_{1}$ automatically based on voxel-size, and we introduce a user-defined multiplier to increase or decrease $t_{1}$, giving the user control over the trade-off between performance and quality that our technique introduces. For more information on the selection of $t_i$, we refer to our supplemental material.

\begin{figure} 
  \begin{tabular}{c c c}
  \textbf{HWRT only} &
  \textbf{Combined (Ours)} &
  \textbf{DF Only} \\
  \includegraphics[trim={24cm 21cm 35cm 1cm},clip,width=0.3\linewidth]{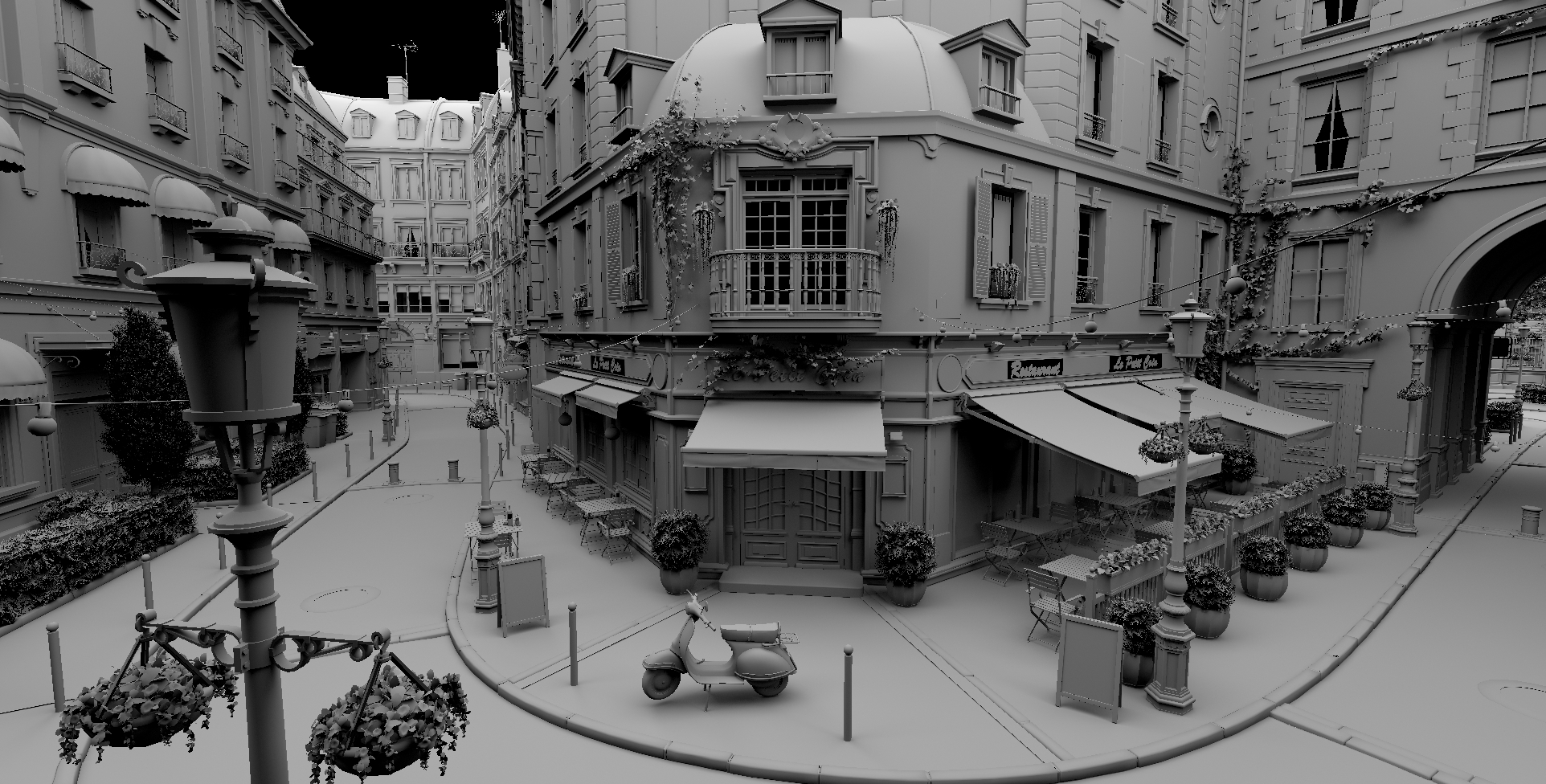} &
  \includegraphics[trim={24cm 21cm 35cm 1cm},clip,width=0.3\linewidth]{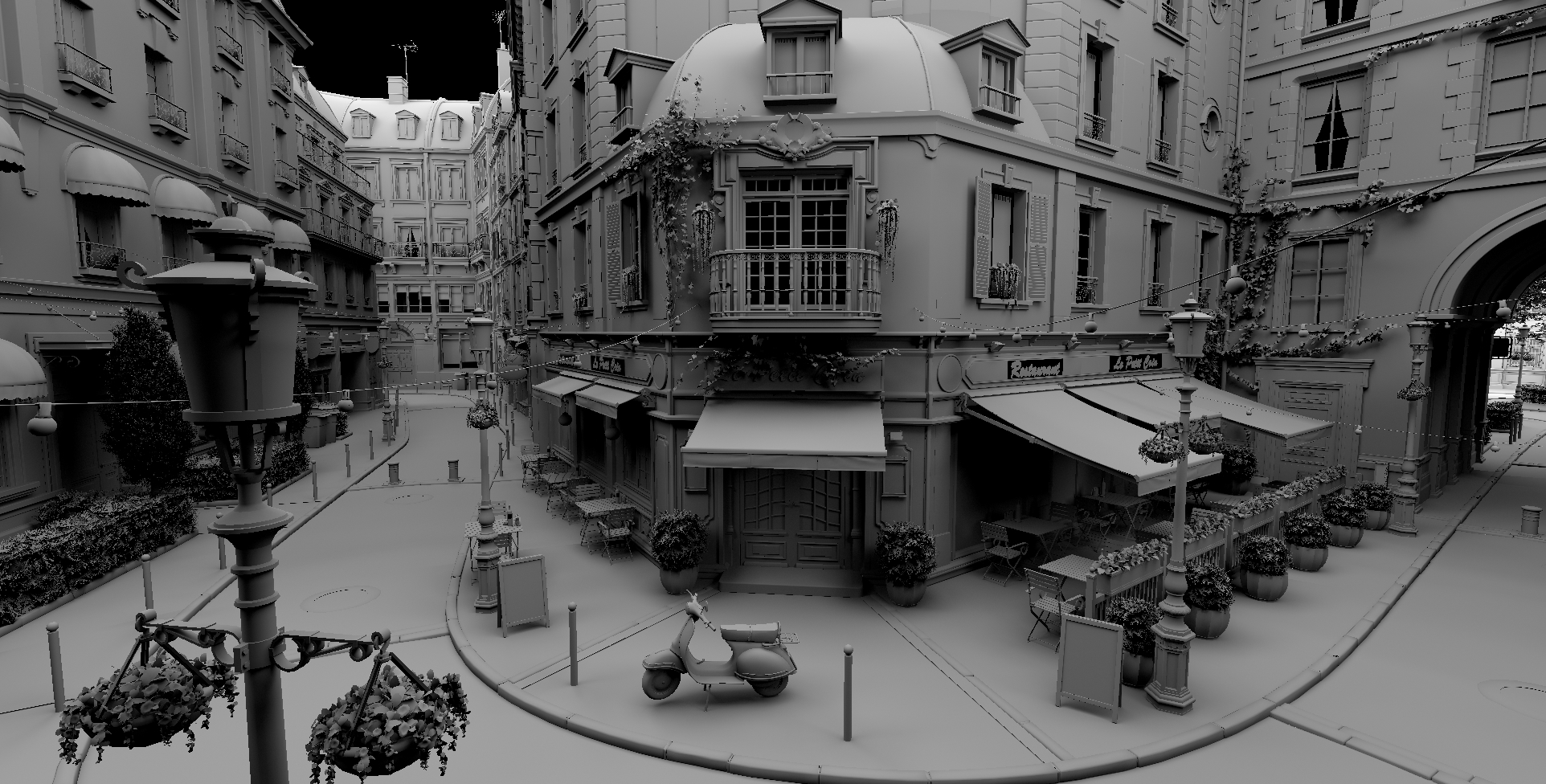} &
  \includegraphics[trim={24cm 21cm 35cm 1cm},clip,width=0.3\linewidth]{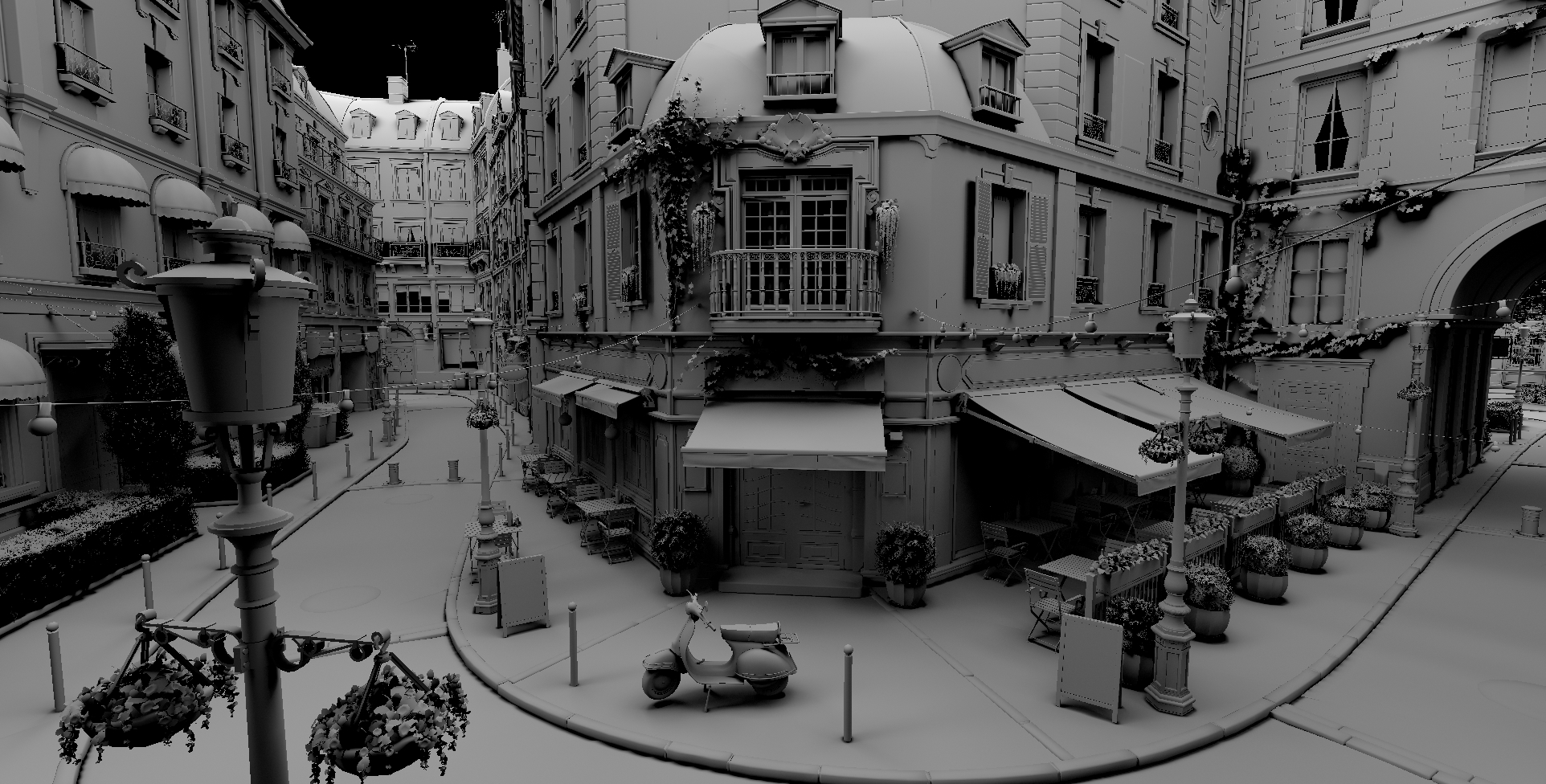}\\
  
  \includegraphics[trim={4cm 8cm 55cm 18cm},clip,width=0.3\linewidth]{fig/gi/1_bvh.jpg} &
  \includegraphics[trim={4cm 8cm 55cm 18cm},clip,width=0.3\linewidth]{fig/gi/1_hybrid.jpg} &
  \includegraphics[trim={4cm 8cm 55cm 18cm},clip,width=0.3\linewidth]{fig/gi/1_brix.jpg}
  \end{tabular}
  \caption{Close-ups of the type of darkening artefacts our method improves in the Bistro scene from Figure \ref{ao} as well in GI-1.0 from Figure \ref{gi10_results}. In the bottom right, we see that the front of the tree is too bright. This is caused by the large normal bias we introduce to avoid self-intersection artefacts. We can also see an incorrect shadow due to \textit{blobbing}.}
  \label{closeup}
\end{figure}

Figure \ref{ao} compares our technique with \textit{pure} tracing on different scenes from various viewpoints. To avoid skewing the results with effects from temporal reprojection or denoising algorithms, we focus on the visual results on static scenes after several frames of accumulation. On a more complex scene such as Bistro, and with long and divergent AO rays, we manage a speed-up of 2.6 in tracing time. Note that this speed-up is dependent on the complexity of the scene, the ray length, and the divergence of the rays. The influence of ray length is shown separately in our supplemental material. When we set a shorter AO distance on simpler scenes (second and third row of Fig. \ref{ao}), we get speed-ups between 1.05 and 1.5. Even for these basic scenarios, the overhead of performing both tracing calls is overcome. An important caveat here is that this speed-up needs to be balanced with the additional overhead of updating the distance field. Nonetheless, we see an advantage is easily achieved on the more complex Bistro scene or for static scenes. In all scenarios, our results are visually much closer to the hardware ray traced reference images, as can be seen in the supplemental material. While some of the overall darkening remains, objectionable artefacts are removed. This can also be seen in more detail in Figure \ref{closeup}. Overall, we find the desired middle ground between the two pure techniques.



\paragraph{GI \& Shadow Rays.} We also validate our technique with shadow rays in the context of Global Illumination, specifically by implementing it within the recently released GI-1.0 Global Illumination pipeline \cite{gi10}. We focus our attention on the two passes of this pipeline that generate hardware ray queries: one pass emits rays to compute indirect illumination and a later pass computes shadow rays to the sampled light sources. The origin of both rays is related to the point seen through the viewport, although both these passes run at a lower resolution than the framebuffer. For the first pass, a disadvantage of our method is that the accuracy of the material information retrieved at the intersection point depends on the tracing technique. In the context of GI-1.0, however, this discrepancy is not problematic, as we are retrieving material information that will be mollified. 
Hence, for our experiment, we make assumptions (i.e. constant albedo), but for best results the distance field could be supplemented with a surface cache.

For the first pass, we use the same division scheme as above: we decide on a single splitting value $t_1$, and trace with HWRT from the origin to this value, and against the distance field after. For the shadow rays in the second pass, we want accurate tracing both at the start of the ray and at the end, near the light source. This is because deformations of the geometry around the light source in the distance field could lead to obvious artefacts in the shadows. Therefore, we define two splitting values, $t_1$ and $t_2$, where we use distance field tracing for the interval $[t_1, t_2]$, and use HWRT for the other two intervals. In this case, we can use any hit tracing. To choose the values of $t_1$ and $t_2$, we again base this on the voxel-size at the origin and end point of the ray, respectively. We again allow the user to alter this distance by way of a user-defined multiplier.

The results of our technique with GI-1.0 are shown in Figure \ref{gi10_results}. The speed-up is lower in this case, since the GI-1.0 algorithm contains various other passes, but we see a solid performance gain nonetheless. Without paying the full cost of HWRT, we again avoid the major visual artefacts stemming from the use of the distance field, which is also shown in more detail in Figure \ref{closeup}.
\section{Conclusion}

We have shown that it is possible to achieve a similar visual quality to hardware ray traced results, with no artefacts and at lower cost, by breaking up a single ray query into two, or even three, separate tracing calls and processing a subset of those traces with distance field tracing. By intelligently choosing where to use hardware ray tracing and where to fall back to distance field tracing, we create a user-controlled middle ground between the advantages of these two techniques that overcomes the associated overhead. We minimize this overhead by decreasing the resolution of the distance field, which is permissible in our combined tracing method as it impacts visual results far less. In general, combined tracing becomes more viable as tracing time starts to dominate, such as when the scene gets more complex, the rays become longer, or more divergent.

\begin{acks}
We would like to thank the reviewers for their helpful comments. We also thank Guillaume Boissé for the voxelization and distance field generation code, and Joel Jordan and H\'elo\"ise Dupont de Dinechin for valuable feedback during the writing process. AMD, AMD Radeon and the AMD Arrow logo, and combinations thereof are trademarks of Advanced Micro Devices, Inc.  Other product names used in this publication are for identification purposes only and may be trademarks of their respective companies.
\end{acks}

\bibliographystyle{ACM-Reference-Format}
\bibliography{main}

\newpage

\end{document}